\documentclass[pre,twocolumn,aps]{revtex4-1}
\usepackage{graphicx}

\usepackage{amsmath,amssymb,bm,amsthm}
\usepackage{braket}

\def\beq{\begin{equation}}
\def\eeq{\end{equation}}
\def\nbeq{\begin{equation*}}
\def\neeq{\end{equation*}}

\def\<{\langle}
\def\>{\rangle}

\def\Tr{{\rm Tr}}

\renewcommand{\d}{\partial}

\newcommand{\balign}[1]{\begin{align} #1 \end{align}}

\newcommand{\todayd}{\the\year/\the\month/\the\day}

\newcommand{\lb}{\label}

\newcommand{\bel}{\begin{easylist}}
\newcommand{\eel}{\end{easylist}}

\def \({\left(}
\def \){\right)}
\def \[{\left[}
\def \]{\right]}

\newcommand{\sumtwo}[2]%
{\mathop{\sum_{#1}}_{#2}}
\newcommand{\sumthree}[3]%
{\mathop{\mathop{\sum_{#1}}_{#2}}_{#3}}
\newcommand{\sumfour}[4]%
{\mathop{\mathop{\mathop{\sum_{#1}}_{#2}}_{#3}}_{#4}} 
\newcommand{\prodtwo}[2]%
{\mathop{\prod_{#1}}_{#2}}
\newcommand{\mintwo}[2]%
{\mathop{\min_{#1}}_{#2}}
\newcommand{\maxtwo}[2]%
{\mathop{\max_{#1}}_{#2}}
\newcommand{\maxthree}[3]%
{\mathop{\mathop{\max_{#1}}_{#2}}_{#3}}
\newcommand{\limtwo}[2]%
{\mathop{\lim_{#1}}_{#2}}
\newcommand{\suptwo}[2]%
{\mathop{\sup_{#1}}_{#2}}
\newcommand{\supthree}[3]%
{\mathop{\mathop{\sup_{#1}}_{#2}}_{#3}}
\newcommand{\supfour}[4]%
{\mathop{\mathop{\mathop{\sup_{#1}}_{#2}}_{#3}}_{#4}} 
\newcommand{\inftwo}[2]%
{\mathop{\inf_{#1}}_{#2}}
\newcommand{\infthree}[3]%
{\mathop{\mathop{\inf_{#1}}_{#2}}_{#3}}
\newcommand{\inffour}[4]%
{\mathop{\mathop{\mathop{\inf_{#1}}_{#2}}_{#3}}_{#4}} 

\newcommand\calQ{{\cal Q}}

\newcommand{\sref}[1]{Sec.~\ref{sec:#1}}

\begin{document}
\title{Thermalization without eigenstate thermalization hypothesis after a quantum quench}

\author{Takashi Mori}
\affiliation{
Department of Physics, Graduate School of Science, University of Tokyo, 
7-3-1 Hongo, Bunkyo-ku, Tokyo 113-0033, Japan
}
\author{Naoto Shiraishi}
\affiliation{
Department of Physics, Keio University, 3-14-1 Hiyoshi, Yokohama 223-8522, Japan
}

\begin{abstract}
Nonequilibrium dynamics of a nonintegrable system without the eigenstate thermalization hypothesis is studied.
It is shown that, in the thermodynamic limit, this model thermalizes after an arbitrary quantum quench at finite temperature, although it does not satisfy the eigenstate thermalization hypothesis.
In contrast, when the system size is finite and the temperature is low enough, the system may not thermalize.
In this case, the steady state is well described by the generalized Gibbs ensemble constructed by using highly nonlocal conserved quantities.
We also show that this model exhibits prethermalization, in which the prethermalized state is characterized by nonthermal energy eigenstates.
\end{abstract}
\maketitle

\section{Introduction}
Approach to thermal equilibrium, or thermalization, in isolated many-body quantum systems has recently attracted renewed interest~\cite{Tasaki1998,Popescu2006,Goldstein2006,Rigol2008,Reimann2008,Goldstein2010,Kinoshita2006,Gring2012}.
Although the quantum state itself does not relax to the (micro)canonical enesmble, isolated systems can thermalize in the sense that the expectation values of any local quantities approach their equilibrium values.
Recent intensive theoretical and experimental studies have revealed that a wide class of far-from-integrable systems with no local conserved quantity thermalize as expected~\cite{Rigol2008, Kinoshita2006, Gring2012}, but systems with some local conserved quantities~\cite{Rigol2007,Biroli2010,Santos2010,Steinigeweg2013,Ilievski2015,Hamazaki2016} and with many-body localization~\cite{Basko2006,Pal2010} do not thermalize.
In addition, it has been recognized that isolated quantum systems also show rich phenomena in the course of relaxation.
For example, some nearly-integrable systems undergo \emph{prethermalization}~\cite{Berges2004,Moeckel2008,Kollar2011,Gring2012,Smith2013,Langen2013,Langen2015,Kitagawa2011,Kaminishi2015}, where relaxation occurs with two-step: 
The first one is to a long-lived prethermalized state and the second one is to the true thermal equilibrium.

What distinguishes whether the system thermalizes?
It is now recognized that the \emph{eigenstate thermalization hypothesis} (ETH), which is traced back to von Neumann~\cite{Neumann1929}, plays an important role to characterize systems with thermalization~\cite{Deutsch1991,Srednicki1994,Rigol2008,Goldstein2015, Palma2015, D'Alessio2016_review}.
The ETH claims that every energy eigenstate is \emph{thermal}, i.e., indistinguishable from the corresponding (micro)canonical ensemble as long as we consider local observables.
By assuming the ETH, we can show thermalization of an isolated quantum system.
Numerical simulations show that many far-from integrable systems with no local conserved quantity satisfy the ETH~\cite{Kim2014,Beugeling2014}, while integrable systems~\cite{Rigol2007,Biroli2010,Santos2010,Steinigeweg2013,Ilievski2015} and localized systems~\cite{Basko2006,Pal2010} do not.
Thus, it looks plausible to consider that the ETH gives a complete characterization of thermalizing systems.

In this paper, contrary to the above scenario, we exemplify that the ETH does \textit{not} fully determine the presence or absence of thermalization by studying the quantum dynamics of a concrete spin model.
We show that, despite the fact that the ETH is \textit{not} satisfied in this model, it thermalizes after any physical quench, i.e., a sudden change of the Hamiltonian, from a finite-temperature equilibrium state with a sufficiently large system size.
Note that our model has only short-range interactions, translation invariance (in particular, no localization), no local conserved quantities (and therefore nonintegrable), but does not satisfy the ETH.
This class of models was first proposed in Ref.~\cite{Shiraishi-Mori2017}, while the nonequilibrium dynamics of such a Hamiltonian has not been studied so far.

On the other hand, when the system size is relatively small and the temperature is low enough, we may find that the system does not thermalize after a quench.
In this case, we numerically show that the steady state is well described by the generalized Gibbs ensemble (GGE) associated with highly {\it nonlocal} conserved quantities.
This shows clear contrast to the case of integrable systems that a non-thermal steady state is described by the GGE characterized by local or quasi-local conserved quantities~\cite{Rigol2007, Sotiriadis2014, Wouters2014, Pozsgay2014, Vidmar2016, Ilievski2015, Essler2016_review}.
The relevance of the GGE constructed by nonlocal conserved quantities brings new insight into the problem of thermalization; some nonlocal conserved quantities can affect the local property of the steady state of an isolated quantum many-body system.

The model studied here also exhibits intriguing nonequilibrium dynamics.
We show that our model undergoes prethermalization for certain initial states.
It turns out that the prethermalized state is characterized by nonthermal energy eigenstates although the true steady state is not affected by them.
It crucially depends on the initial state whether the system thermalizes directly or thermalizes via prethermalization plateau, in contrast to the conventional prethermalization in nearly integrable systems, where any initial state shows prethermalization unless the GGE associated with the initial state is identical to the canonical ensemble.

This paper is organized as follows.
In \sref{model}, we explain the setup of our model, which is constructed by the method of embedded Hamiltonian.
In \sref{thermalization}, we confirm analytically and numerically that a macroscopic system thermalizes without the ETH after a physical quench.
By contrast, in \sref{no-therm} we show that a sufficiently small system may not thermalize.
In this case, the stationary state is described by a novel type of generalized Gibbs ensemble characterized by highly non-local observables.
In \sref{pre}, we investigate the signature of non-local observables in macroscopic systems.
We find that a novel type of prethermalization occurs in the course of relaxation.

\section{Model}\lb{sec:model}
We consider a one-dimensional spin-1 chain of length $L$ under the periodic boundary condition.
Each site is labeled as $i=1,2,\dots,L$.
The spin-1 operator is denoted by $\hat{\bm{S}}$ (the spin-1 operator at site $i$ is denoted by $\hat{\bm{S}}_i$), and the three eigenstates of $\hat{S}^z$ are denoted by $|1\>$, $|0\>$, and $\ket{-1}$, where $\hat{S}^z|m\>=m|m\>$ for $m=-1,0,1$.
We define $\hat{P}:=|0\>\<0|$ as the projection operator to the state $|0\>$, and denote by $\hat{P}_i$ the operator acting on the site $i$.
We also introduce the pseudo Pauli matrix $\tilde{\bm{\sigma}}$ as
\beq
\left\{
\begin{split}
\tilde{\sigma}^x&=|1\>\<-1|+|-1\>\<1|, \\
\tilde{\sigma}^y&=-i|1\>\<-1|+i|-1\>\<1|, \\
\tilde{\sigma}^z&=|1\>\<1|-|-1\>\<-1|.
\end{split}
\right.
\eeq

The Hamiltonian studied in this paper is given by
\beq
\hat{H}=\sum_{i=1}^L\hat{h}_i\hat{P}_i+\hat{H}' \equiv \hat{H}_0+\hat{H}',
\label{eq:H}
\eeq
where $\hat{1}$ represents identity operator, and we set
\balign{
\hat{h}_i&=\sum_{\alpha=x,y,z}\left[J_{\alpha}\hat{S}_{i-1}^{\alpha}\hat{S}_{i+1}^{\alpha}-h_{\alpha}(\hat{S}_{i-1}^{\alpha}+\hat{S}_{i+1}^{\alpha})\right]+D\hat{1},  \\
\hat{H}'&=\sum_{i=1}^L\sum_{\alpha=x,y,z}\left(J_{\alpha}'\tilde{\sigma}_i^{\alpha}\tilde{\sigma}_{i+1}^{\alpha}-h_{\alpha}'\tilde{\sigma}_i^{\alpha}\right).
}
In this paper, the parameters are fixed at $\vec{J}=(-0.75,0.5,1)$, $\vec{h}=(1,0,-0.5)$, $\vec{J'}=(-1.5,1,-2)$, $\vec{h'}=(0,0,-0.5)$, and $D=-1$.
This Hamiltonian belongs to a class of embedded Hamiltonians studied in Ref.~\cite{Shiraishi-Mori2017}.

We introduce the Hilbert subspace $\mathcal{T}\subset\mathcal{H}$ defined as the collection of the states $|\Psi\>\in\mathcal{H}$ satisfying $\hat{P}_i|\Psi\>=0$ for all $i$.
In our model, the dimension of the whole Hilbert space $\mathcal{H}$ and that of its subspace $\mathcal{T}$ are $\mathrm{dim}\,\mathcal{H}=3^L$ and $\mathrm{dim}\,\mathcal{T}=2^L$, respectively.
Owing to $[\hat{H}',\hat{P}_i]=0$ for all $i$, the set of $2^L$ eigenstates of $\hat{H}'$ serves as the orthonormal basis of $\mathcal{T}$.
In addition, these $2^L$ eigenstates are also energy eigenstates of $\hat{H}$ because $\hat{H}_0|\Psi\>=0$ for any $|\Psi\>\in\mathcal{T}$.
Thus the eigenstates of $\hat{H}'$ within $\mathcal{T}$ are embedded to the energy spectrum of a more complicated Hamiltonian $\hat{H}=\hat{H}_0+\hat{H}'$.
It is proved that these $2^L$ energy eigenstates within $\mathcal{T}$ are not thermal, and $\hat{H}$ does not satisfy the ETH~\cite{Shiraishi-Mori2017}.

It is noted that, for general choices of the parameters $\{J_{\alpha},h_{\alpha},J_{\alpha}',h_{\alpha}'\}$, $\hat{H}$ does not have any local conserved quantities.
The Hamiltonian $\hat{H}$ is therefore an example of translation invariant local Hamiltonians with no local conserved quantities which do not satisfy the ETH.

For later convenience, we introduce the projection operator onto the subspace $\mathcal{T}$ defined as
\beq
\hat{\mathcal{Q}}:=\prod_{i=1}^L(1-\hat{P}_i)=\prod_{i=1}^L\hat{Q}_i,
\eeq
where $\hat{Q}_i:=1-\hat{P}_i$.
This projection operator $\hat{\mathcal{Q}}$ is a conserved quantity of our model.
This is a highly nonlocal quantity in the sense that it is written as a product of local projection operators at every site $i$.

\section{Thermalization}
\label{sec:thermalization}

\subsection{Preliminary discussion on the ETH and the weak ETH}

Our model has $2^L(=\dim\mathcal{T})$ nonthermal energy eigenstates.
Since the dimension of the total Hilbert space $\mathcal{H}$ is $3^L$, the fraction of nonthermal energy eigenstates,
\beq
\frac{\dim\mathcal{T}}{\dim\mathcal{H}}=\left(\frac{2}{3}\right)^L
\eeq
is exponentially small for large system sizes.

When \textit{almost all} the energy eigenstates of the system are thermal, the system is said to satisfy the \textit{weak ETH}~\cite{Biroli2010} (here, almost all means that the fraction of nonthermal energy eigenstates tends to zero in the thermodynamic limit).
Our model does not satisfy the ETH, but the weak ETH holds.

One might think that the statement of the ETH (\textit{all} the energy eigenstates should be thermal) is too strong, and the weak ETH is a more important property to determine the presence or absence of thermalization.
However, this is not the case.
It is proved that the weak ETH holds for general translation-invariant quantum spin chains with local interactions, and moreover, the fraction of nonthermal energy eigenstates is exponentially small for large system sizes~\cite{Mori2016_weak}.
In other words, even integrable systems satisfy the weak ETH, although they fail to thermalize after a quench.
In the integrable case, nonthermal energy eigenstates with exponentially small fraction have extremely large weight after some physical quench~\cite{Biroli2010}, which causes absence of thermalization.
Thus, even if the number of nonthermal energy eigenstates is exponentially fewer than the total number of energy eigenstates, it is highly nontrivial whether thermalization occurs.

Rather, it is sometimes argued that the ETH is even necessary for thermalization although it looks too strong~\cite{Palma2015}.
It is true that the violation of the ETH implies that there are some specific initial states which do not thermalize.
The problem is whether such initial states are realizable in practice.
In Ref.~\cite{Palma2015}, it is argued that the ETH is necessary for thermalization in the sense that the ETH must hold whenever all product states between a small subsystem and the remaining part called a ``bath'' thermalize.
However, we should be care about the class of initial states; not all the product states might be realizable in experiment, and therefore it would be an excessive requirement that all the product states should thermalize.
In the next subsection, we will see that the ETH is \textit{not} necessary for thermalization in the sense that our system thermalizes after any quantum quench at a finite temperature although the ETH does not hold.

\subsection{Thermalization after any finite-temperature quench}
Now we consider the quantum dynamics after the quench of the Hamiltonian.
We consider the following pre-quench Hamiltonian:
\beq
\hat{H}_{\mathrm{ini}}=\hat{H}-h_x^{(0)}\sum_{i=1}^L\hat{S}_i^x+D^{(0)}\sum_{i=1}^L\hat{P}_i.
\label{eq:Hini}
\eeq
The initial state $|\Psi_{\mathrm{ini}}\>$ is chosen as a canonical thermal pure quantum (TPQ) state~\cite{Sugiura-Shimizu2013} of $\hat{H}_{\mathrm{ini}}$ at the inverse temperature $\beta>0$,
\beq
|\Psi_{\mathrm{ini}}\>=\frac{e^{-\beta\hat{H}_{\mathrm{ini}}/2}|r\>}{\sqrt{\<r|e^{-\beta\hat{H}_{\mathrm{ini}}}|r\>}},
\label{eq:TPQ}
\eeq
where $|r\>$ with $\<r|r\>=1$ is a random vector sampled uniformly from the entire Hilbert space $\mathcal{H}$.
The canonical TPQ state is known to represent thermal equilibrium at the inverse temperature $\beta$~\cite{Sugiura-Shimizu2013}.
After the quench, the system evolves under the Hamiltonian $\hat{H}$, and the state at time $t>0$ is given by $|\Psi(t)\>=e^{-i\hat{H}t}|\Psi_{\mathrm{ini}}\>$, where we put $\hbar=1$.

In Ref.~\cite{Shiraishi-Mori2017}, it is numerically shown that all energy eigenstates of $\hat{H}$ outside of $\mathcal{T}$, i.e., energy eigenstates $\{|\Phi_n\>\}$ with $\<\Phi_n|\hat{\mathcal{Q}}|\Phi_n\>=0$, are thermal.
From this numerical result, we can conclude that the system will thermalize if the initial state has a vanishingly small weight to the nonthermal energy eigenstates in $\mathcal{T}$.
This weight $w$ is equal to the expectation value of $\hat{\mathcal{Q}}$; $w=\<\Psi_{\mathrm{ini}}|\hat{\mathcal{Q}}|\Psi_{\mathrm{ini}}\>$.

Now we shall prove that the initial state \eqref{eq:TPQ} actually has an exponentially small weight to the subspace $\mathcal{T}$.
First, we consider the quantity
\beq
\hat{A}:=\frac{1}{L}\sum_{i=1}^L\hat{P}_i.
\eeq
It is shown in Ref.~\cite{Shiraishi-Mori2017} that the canonical distribution with $\hat{H}_{\rm ini}$ has finite expectation value of $\hat{A}$
\beq
\<\hat{A}\>_{\mathrm{can}}^{(\mathrm{ini})}=\frac{1}{L}\sum_{i=1}^L\<\hat{P}_i\>_{\mathrm{can}}^{(\mathrm{ini})}\geq c_{\beta},
\label{eq:P_can}
\eeq
where $\langle\hat{O}\rangle_{\mathrm{can}}^{\mathrm{(ini)}}:=\mathrm{Tr}\,\hat{O}e^{-\beta\hat{H}_{\mathrm{ini}}}/\mathrm{Tr}\,e^{-\beta\hat{H}_{\mathrm{ini}}}$ is the average in the canonical distribution, and $c_{\beta}$ is a strictly positive quantity independent of the system size $L$ for any finite temperature.
It is noted that any state in $\mathcal{T}$ has exactly zero expectation value of $\hat{A}$ and vise versa, which imply that $\hat{\mathcal{Q}}$ is identical to the projection operator onto the subspace spanned by the eigenstates of $\hat{A}$ with the zero eigenvalue, which is denoted by $\hat{\mathsf{P}}[\hat{A}=0]$:
\beq
\hat{\mathcal{Q}}=\hat{\mathsf{P}}[\hat{A}=0].
\label{eq:Q_A}
\eeq

Let $\hat{B}$ be an arbitrary self-adjoint operator, and let $b_k$ and $|b_k\>$ ($k=1,2,\dots$) be the corresponding eigenvalues and eigenstates, respectively.
We denote by
\beq
\hat{\mathsf{P}}[\hat{B}\geq b]:=\sum_{k:b_k\geq b}|b_k\>\<b_k|
\eeq
the projection operator onto the subspace spanned by the eigenstates of $\hat{B}$ with eigenvalues not less than $b$.
Then, owing to the large deviation property of the equilibrium state in quantum spin chains, we can prove that for an arbitrary $\delta>0$ and for sufficiently large $L_0$, there exists $\gamma>0$ such that
\beq
\left\<\hat{\mathsf{P}}\left[\left|\hat{A}-\<\hat{A}\>_{\mathrm{can}}^{\mathrm{(ini)}}\right|\geq\delta\right]\right\>_{\mathrm{can}}^{\mathrm{(ini)}}
\leq e^{-\gamma L}
\label{eq:LDP}
\eeq
holds for any $L\geq L_0$~\cite{Ogata2010,Tasaki2016}.
From Eqs.~(\ref{eq:P_can}) and (\ref{eq:Q_A}), it is obvious that
\beq
\<\hat{\mathcal{Q}}\>_{\mathrm{can}}^{\mathrm{(ini)}}\leq\left\<\hat{\mathsf{P}}\left[\left|\hat{A}-\<\hat{A}\>_{\mathrm{can}}^{\mathrm{(ini)}}\right|\geq\delta\right]\right\>_{\mathrm{can}}^{\mathrm{(ini)}}
\eeq
if we choose $\delta$ so that $0<\delta<c_{\beta}$.
By using Eq.~(\ref{eq:LDP}), we obtain
\beq
\<\hat{\mathcal{Q}}\>_{\mathrm{can}}^{(\mathrm{ini})}\leq e^{-\gamma L}.
\eeq
The expectation value of $\hat{\mathcal{Q}}$, i.e. the weight to the subspace $\mathcal{T}$, in the initial canonical distribution is exponentially small.

Since a canonical TPQ state represents thermal equilibrium, it would be expected that the weight to the subspace $\mathcal{T}$ is also exponentially small for almost all realizations of the canonical TPQ state.
Indeed, we can prove the following inequality, which rigorously establishes that the weight to the nonthermal energy eigenstates decreases exponentially as the system size increases:
\beq
\mathrm{Prob}\left[\<\Psi_{\mathrm{ini}}|\hat{\mathcal{Q}}|\Psi_{\mathrm{ini}}\>\geq e^{-\gamma'L}\right]\leq (4+2\sqrt{2})e^{-\gamma'L},
\label{eq:prob_weight}
\eeq
where $\gamma'>0$ and $\mathrm{Prob}[a]$ denotes the probability of an event $a$, and the probability is introduced for a random vector construction of the canonical TPQ state~(\ref{eq:TPQ}).
The proof of Eq.~(\ref{eq:prob_weight}) is given in Appendix.~\ref{app:TPQ}.
Although we have focused on the canonical TPQ state, it is expected that the weight also decreases exponentially for other realistic initial states.

\begin{figure}[t]
\begin{center}
\includegraphics[clip,width=70mm]{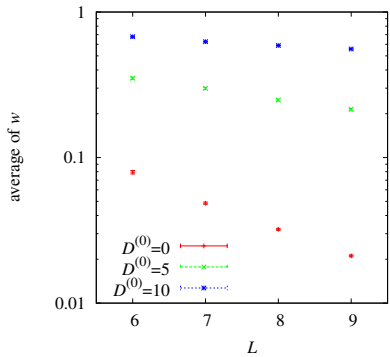}
\caption{Exponential decrease of the average weight to nonthermal energy eigenstates as a function of $L$ with error bars.
The average is taken over 100 realizations of the random vector for the canonical TPQ state with $\beta=0.2$.}
\label{fig:weight}
\end{center}
\end{figure}

\begin{figure}[t]
\begin{center}
\begin{tabular}{c}
\includegraphics[clip,width=70mm]{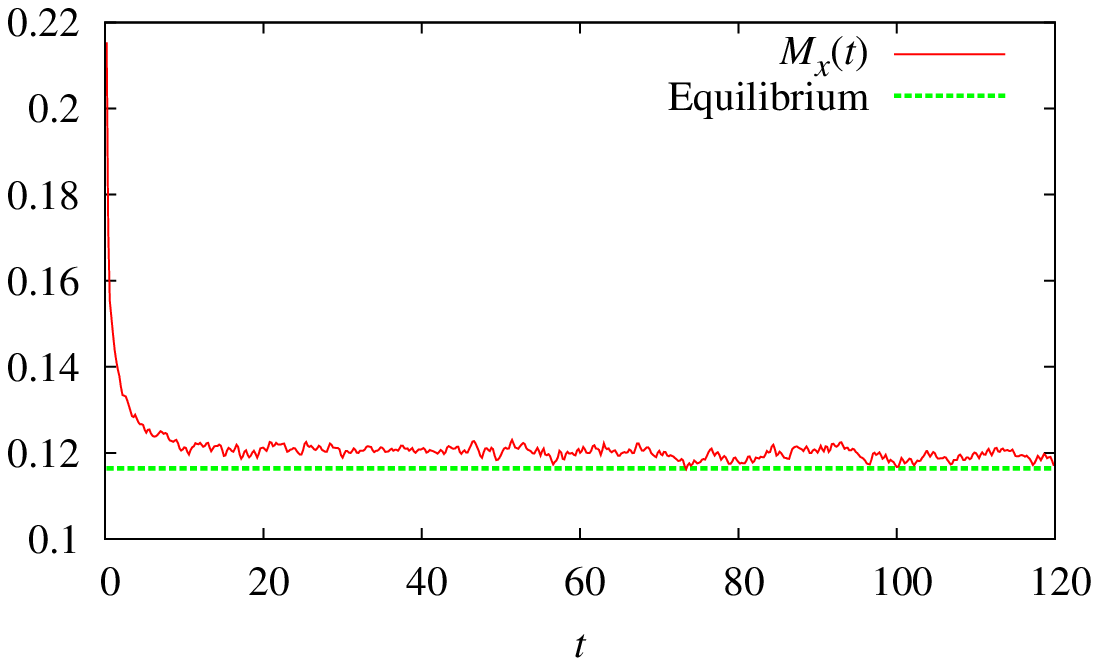}\\
\includegraphics[clip,width=70mm]{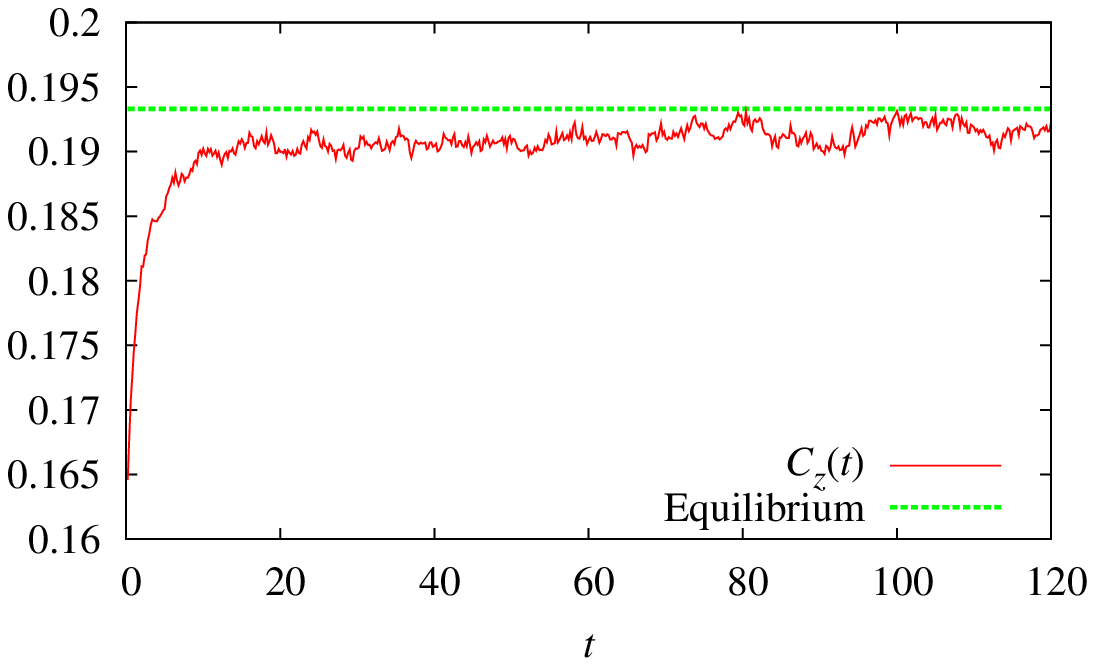}
\end{tabular}
\caption{Time evolutions of $M^x(t)$ (top) and $C_z(t)$ (bottom) for the system size $L=12$ after a quantum quench to the Hamiltonian \eqref{eq:H} from a thermal state of the Hamiltonian \eqref{eq:Hini} with $D^{(0)}=0$, $h_x^{(0)}=0.5$, and $\beta=0.2$. 
These quantities relax to the corresponding equilibrium values.}
\label{fig:thermal}
\end{center}
\end{figure}

We numerically demonstrate thermalization after the quench.
The parameters in the pre-quench Hamiltonian~(\ref{eq:Hini}) are set to $h_x^{(0)}=0.5$ and $D^{(0)}=0$, and the inverse temperature before the quench is set as $\beta=0.2$.
The numerical results for the average weight to the nonthermal energy eigenstates as a function of $L$ are depicted as red points in FIG.~\ref{fig:weight}.
The average is taken over 100 realizations of random vectors constructing canonical TPQ states.
As clearly seen from FIG.~\ref{fig:weight}, the exponential decrease of $w$ is numerically confirmed.
Time evolutions of macroscopic observables also confirm the presence of thermalization.
The typical time evolutions of $M_x(t):=(1/L)\sum_{i=1}^L\<\Psi(t)|S_i^x|\Psi(t)\>$ and $C_z(t):=(1/L)\sum_{i=1}^L\<\Psi(t)|S_i^zS_{i+1}^z|\Psi(t)\>$ depicted in FIG.~\ref{fig:thermal} exhibit thermalization to the corresponding equilibrium values.

\section{No thermalization in finite systems}
\begin{figure}[t]
\begin{center}
\begin{tabular}{c}
\includegraphics[clip,width=70mm]{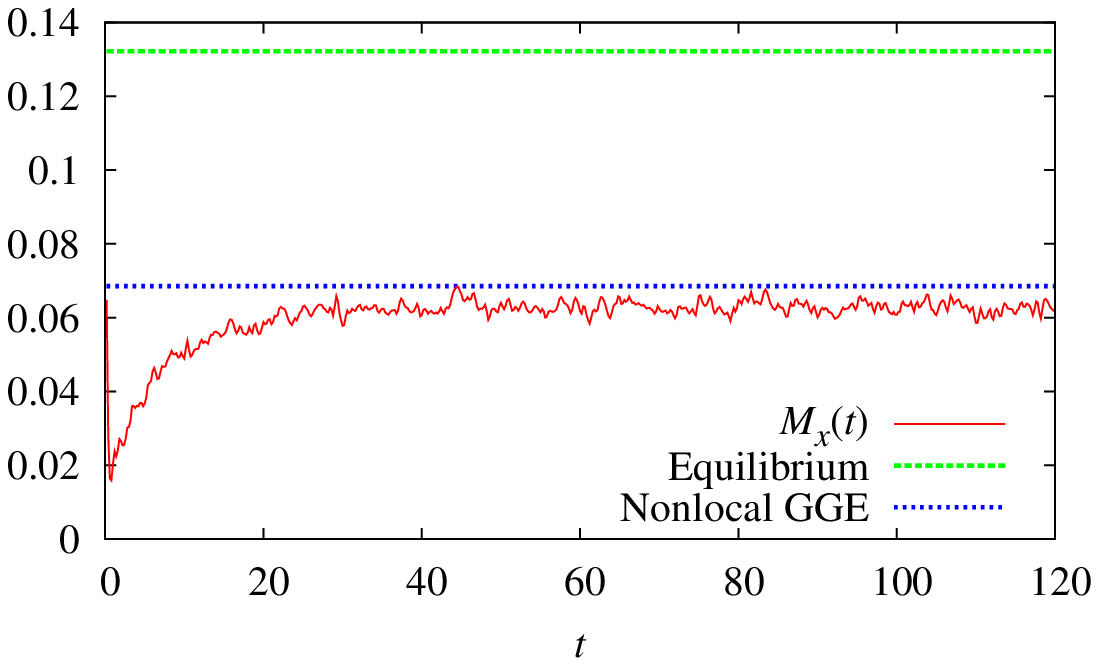}\\
\includegraphics[clip,width=70mm]{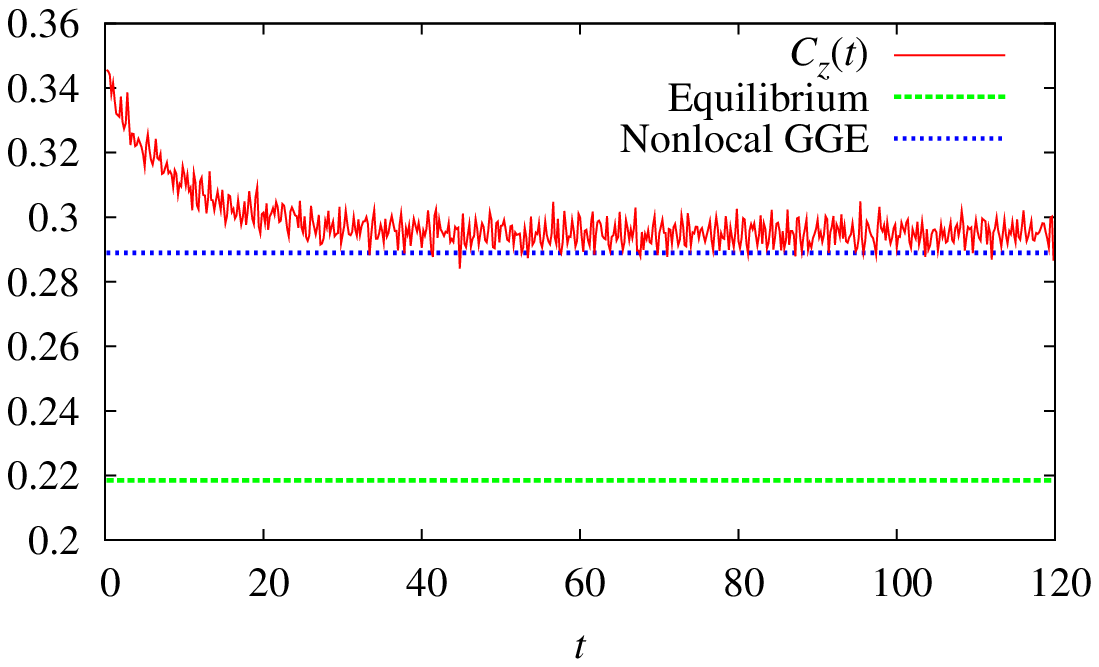}
\end{tabular}
\caption{Time evolutions of $M_x(t)$ (top) and $C_z(t)$ (bottom) for $L=12$ after a quantum quench to the Hamiltonian \eqref{eq:H} from a thermal state of the Hamiltonian \eqref{eq:Hini} with $D^{(0)}=10$, $h_x^{(0)}=0.5$, and $\beta=0.2$.
These quantities relax not to their equilibrium values, but to their values predicted by the nonlocal GGE.}
\label{fig:GGE}
\end{center}
\end{figure}

\lb{sec:no-therm}
In contrast to the previous section, if the system size is finite (i.e., finite $L$), the weight $w$ can be relatively large, and the system will not thermalize in that case.
We consider the same quench, but put $D^{(0)}=10$.
A larger value of $D^{(0)}$ in the pre-quench Hamiltonian will lead to larger $w$ (the expectation value of $\hat{\mathcal{Q}}$), although for any $D^{(0)}$ $w$ decreases exponentially for large $L$, see FIG.~\ref{fig:weight}.
The typical time evolutions of $M_x(t)$ and $C_z(t)$ for $L=12$ are shown in FIG.~\ref{fig:GGE}.
The system approaches a steady state, but it is different from the equilibrium state.

What is this nonthermal steady state?
It is known that a nonthermal steady state of an integrable system is described by the generalized Gibbs ensemble (GGE), $\rho\propto\exp[-\sum_{\alpha}\lambda_{\alpha}\hat{I}_{\alpha}]$, where $\{\hat{I}_{\alpha}\}$ are local or quasi-local conserved quantities of the integrable system, and $\{\lambda_{\alpha}\}$ are determined by the expectation values of those conserved quantities in the initial state~\cite{Rigol2007,Sotiriadis2014,Ilievski2015,Essler2016_review}.
By contrast, our Hamiltonian $\hat{H}$ has no local conserved quantity.
Instead, we have a nonlocal conserved quantity $\hat{\mathcal{Q}}$.
In addition, the Hamiltonian $\hat{H}$ is decomposed as $\hat{H}=\hat{\mathcal{P}}\hat{H}\hat{\mathcal{P}}+\hat{\mathcal{Q}}\hat{H}\hat{\mathcal{Q}}$, where $\hat{\mathcal{P}}:=1-\hat{\mathcal{Q}}$, and then both of $\hat{\mathcal{P}}\hat{H}\hat{\mathcal{P}}$ and $\hat{\mathcal{Q}}\hat{H}\hat{\mathcal{Q}}$ are also nonlocal conserved quantities.
By using them, let us construct the \textit{nonlocal GGE} as
\beq
\hat{\rho}_{\mathrm{GGE}}=\frac{1}{Z_{\mathrm{GGE}}}e^{-\beta_P\hat{\mathcal{P}}\hat{H}\hat{\mathcal{P}} -\beta_Q\hat{\mathcal{Q}}\hat{H}\hat{\mathcal{Q}}-\lambda\hat{\mathcal{Q}}},
\label{eq:nonlocal_GGE}
\eeq
where $Z_{\mathrm{GGE}}$ is determined so that $\mathrm{Tr}\,\rho_{\mathrm{GGE}}=1$.
The parameters $\beta_P$, $\beta_Q$, and $\lambda$ are determined by the conditions $\mathrm{Tr}\, [\hat{\mathcal{O}}\rho_{\mathrm{GGE}}]=\<\Psi_{\mathrm{ini}}|\hat{\mathcal{O}}|\Psi_{\mathrm{ini}}\>$ for $\hat{\mathcal{O}}=\hat{\mathcal{P}}\hat{H}\hat{\mathcal{P}}, \hat{\mathcal{Q}}\hat{H}\hat{\mathcal{Q}}$, and $\hat{\mathcal{Q}}$.

We compare $M_x(t)$ and $C_z(t)$ with the corresponding expectation values in the nonlocal GGE in FIG.~\ref{fig:GGE}.
This figure shows that when the system does not thermalize, the steady state is well described by the nonlocal GGE.
It is a novel property of embedded Hamiltonians that highly nonlocal conserved quantities affect the local property of the steady state.

\section{Prethermalization}\lb{sec:pre}
While the weight to the subspace $\mathcal{T}$ is negligibly small for large systems, it does not mean that the nonlocal conserved quantity $\hat{\calQ}$ and the corresponding subspace $\mathcal{T}$ do not play any role in the thermalization process.
Let us call the site $i$ with $\hat{S}_i^z=0$ (or $\hat{P}_i=1$) a \textit{defect}.
We show that if the density of the defects in the initial state, defined as $d_0:=(1/L)\sum_{i=1}^L\<\Psi_{\mathrm{ini}}|\hat{P}_i|\Psi_{\mathrm{ini}}\>$, is sufficiently small, the system exhibits prethermalization, and the prethermalized state is well described by the constrained (micro)canonical ensemble within the subspace $\mathcal{T}$.

The occurrence of prethermalization can be explained as follows (see Appendix~\ref{app:occurrence} for details).
If the initial density of defects $d_0$ is exactly zero, it implies $w=1$ and the system does not thermalize but equilibrates to a state described by the canonical distribution restricted to the energy eigenstates in the subspace $\mathcal{T}$.
In this case, $\hat{H}_0$ in Eq.~(\ref{eq:H}) does not play any role.
Let us consider the case that $d_0$ is nonzero but sufficiently small.
In this case, since the spread of the defects is not quick, it is expected that up to some finite time the density of defects remains very small and we can neglect the effect of $\hat{H}_0$.
Then, this finite time dynamics for sufficiently small $d_0$ will be very close to that for $d_0=0$.
Therefore, the system will first relax to an ``equilibrium state'' within $\mathcal{T}$, and then reach the true equilibrium state.

The lower bound of the length of time of the prethermalization plateau can be obtained by using the Lieb-Robinson bound~\cite{Lieb1972,Hastings2006}, which reads $t_{\mathrm{pre}}^{\mathrm{(LR)}}\sim 1/(d_0v_{\mathrm{LR}})$, where $v_{\mathrm{LR}}$ is the Lieb-Robinson velocity.
The detailed analysis is given in Appendix~\ref{app:occurrence}, and we here give a rough argument in the following.
The average distance of the nearest defects in the initial state is given by $1/d_0$.
According to the Lieb-Robinson bound, the speed of the spread of each defect is rigorously bounded above by the Lieb-Robinson velocity $v_{\mathrm{LR}}$, and as a result, at time $t\ll t_{\mathrm{pre}}^{\mathrm{(LR)}}\sim 1/(d_0v_{\mathrm{LR}})$, defects have not spread over the entire system yet and we can ignore the effect of defects.
We here remark that the obtained bound by using the Lieb-Robinson bound is a lower bound, and as shown below the real dynamics of the defects is diffusive, not ballistic (i.e., scaled as $1/d_0^2$, not $1/d_0$).

\begin{figure}[t]
\begin{center}
\includegraphics[clip,width=70mm]{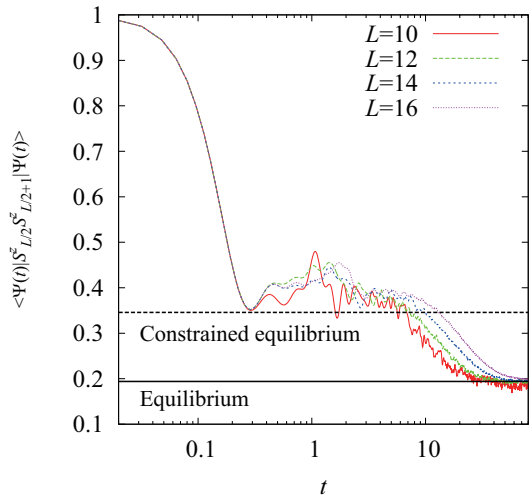}
\caption{The relaxation dynamics of $\<\Psi(t)|S_{L/2}^zS_{L/2+1}^z|\Psi(t)\>$ from $|\Psi_{\mathrm{ini}}\>=|0\>\otimes|1\>\otimes\dots\otimes|1\>$.
It shows thermalization after a prethermalization plateau.
The dashed line indicates the expectation value in the constrained equilibrium state within the subspace $\mathcal{T}$, and the solid line indicates the equilibrium value.}
\label{fig:pre}
\end{center}
\end{figure}

We now study the prethermalization by numerical simulations.
We consider the dynamics starting from the initial state in which the spin at $i=1$ is in the state $|0\>$, and the other spins are in the state $|1\>$: 
\beq
|\Psi_{\mathrm{ini}}\>=|0\>\otimes|1\>\otimes\dots\otimes|1\>.
\eeq
The site $i=1$ is the initial defect, and $d_0=1/L$.
It is noted that even a single defect leads to $w=0$, which implies that the system will eventually thermalize.
We calculate the time evolution of $\<\Psi(t)|\hat{S}_{L/2}^z\hat{S}_{L/2+1}^z|\Psi(t)\>$ (here we assume an even $L$).

The numerical results in FIG.~\ref{fig:pre} show that the prethermalization indeed takes place.
We also find that the prethermalization plateau is well described by the canonical ensemble restricted to $\mathcal{T}$,
\beq
\rho_Q=\frac{\hat{\mathcal{Q}}e^{-\beta_Q\hat{\mathcal{Q}}\hat{H}\hat{\mathcal{Q}}}}{\Tr\[ \hat{\mathcal{Q}}e^{-\beta_Q\hat{\mathcal{Q}}\hat{H}\hat{\mathcal{Q}}}\]}=\frac{\hat{\mathcal{Q}}e^{-\beta_Q\hat{H}}}{\Tr\[ \hat{\mathcal{Q}}e^{-\beta_Q\hat{H}}\]},
\eeq
which is obtained by taking the limit of $\lambda\rightarrow-\infty$ in the nonlocal GGE given in Eq.~(\ref{eq:nonlocal_GGE}).
The effective inverse temperature $\beta_Q$ has been determined so that the expectation value of $\hat{\mathcal{Q}}\hat{H}\hat{\mathcal{Q}}$ in the initial state coincides with that in $\rho_Q$.
We call this state ``constrained equilibrium'', and the average of $\hat{S}_{L/2}^z\hat{S}_{L/2+1}^z$ in this state is indicated by the dashed line in FIG.~\ref{fig:pre}.

In this way, it turns out that although the weight to the subspace $\mathcal{T}$ is exactly zero in our initial state, the system first relaxes to the prethermalized state indistinguishable from thermal equilibrium restricted to $\mathcal{T}$.
This phenomenon shows intriguing nature of the nonequilibrium dynamics of a many-body system.

\begin{figure}[tb]
\begin{center}
\includegraphics[clip,width=100mm]{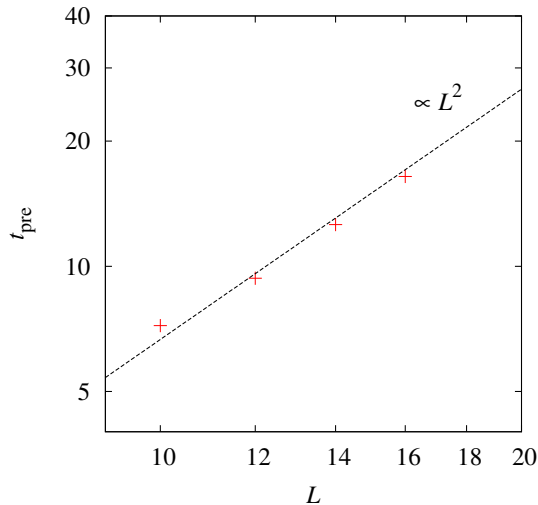}
\caption{The timescale of prethermalization is proportional to $1/d_0^2=L^2$ for the initial state with a single defect.}
\label{fig:lifetime}
\end{center}
\end{figure}

Regarding the lifetime of the prethermalized state, it is numerically found that it scales as 
\beq
t_{\mathrm{pre}}\propto \frac{1}{d_0^2},
\eeq
which suggests diffusive spread of the defects.
In our initial state, $d_0=1/L$ and hence $t_{\mathrm{pre}}\propto L^2$, and this dependence is clearly shown in FIG.~\ref{fig:lifetime}.
In numerical calculations, we have defined $t_{\mathrm{pre}}$ as the time at which $\<\Psi(t)|\hat{S}^z_{L/2}\hat{S}^z_{L/2+1}|\Psi(t)\>$ reaches 0.32.

It should be noted here that the timescale of prethermalization depends on the initial state, and therefore the system does not show prethermalization when $d_0$ is not small in the initial state.
This shows clear contrast to the prethermalization in nearly-integrable systems, which is irrelevant to the initial state.

\section{Summary and Discussion}
We have studied the relaxation process of a spin-1 chain which belongs to the class of embedded Hamiltonians.
It is shown that this system thermalizes after a quench from a thermal state of another local Hamiltonian although the ETH is not satisfied.
However, in a finite system, the weight to nonthermal energy eigenstates may be relatively large, and the system may not thermalize.
We find that, in such a case, the nonlocal GGE associated with some nonlocal conserved quantities describes the steady state.
This means that some highly nonlocal conserved quantities can be relevant to determine the steady state of an isolated quantum system.

In addition, we find that our system exhibits prethermalization described by a constrained equilibrium state within the Hilbert subspace of nonthermal energy eigenstates even if the weight to this subspace is exactly zero.
This result implies that the nonlocal conserved quantities may result in not only the existence of nonthermal energy eigenstate but also the existence of an intermediate quasi stationary state.

It is worth noting the similarity to the glassy dynamics.
A quantum version of the kinetically constraint models, a model of glassy dynamics, also show two-step relaxation which has initial-state dependence~\cite{van2015, Hickey2016, Lan2017}.
In fact, these models can be understood as a special case of the embedded Hamiltonian.
It would be interesting if non-local observables and embedded Hamiltonians characterize glassy dynamics.

Our result elucidates the crucial difference between a quench from a thermal state with finite temperature and that from a ground state (absolute zero temperature).
Since any real experiment is performed at finite temperature, we conclude that the former is a physically realizable quench but the latter is not.
In our model in the thermodynamic limit, we demonstrated that an initial state obtained through the former quench thermalizes, while it is also true that there exists a latter type of quench which provides an initial state without thermalization.
This clearly shows that a naive application of a quench from a ground state may lead to unphysical results.

The reason why our model thermalizes despite no ETH would be understood by the fact that the steady state is described by the nonlocal GGE.
The nonlocal GGE deviates from the equilibrium ensemble only when the expectation values of the relevant nonlocal conserved quantities differ from their equilibrium values.
However, it would be very difficult (or impossible) to control the expectation values of nonlocal quantities by any local operation like a quench at finite temperature.
It means that the expectation values of the nonlocal conserved quantities take the equilibrium values from the beginning, and the nonlocal GGE is reduced to the usual equilibrium ensemble for preparable initial states.
Thermalization without the ETH and the existence of relevant non-local conserved quantities are novel features of our model, and their implications for our understanding of thermalization should be explored further in future works.

\begin{acknowledgments}
TM is supported by  JSPS KAKENHI Grant No. 15K17718, and NS is supported by Grant-in-Aid for JSPS Fellows JP17J00393.
\end{acknowledgments}

\appendix
\section{Exponential decrease of the weight to nonthermal energy eigenstates}
\label{app:TPQ}

We consider a thermal pure quantum state of the Hamiltonian $\hat{H}_{\mathrm{ini}}$ with $L$ spins as~\cite{Sugiura-Shimizu2013}
\beq
|\Psi\>:=\frac{e^{-\beta\hat{H}_{\mathrm{ini}}/2}|r\>}{\sqrt{\<r|e^{-\beta\hat{H}_{\mathrm{ini}}}|r\>}},
\eeq
where $|r\>$ is a random vector defined by using the energy eigenstate $|\phi_n\>$ ($\hat{H}_{\mathrm{ini}}|\phi_n\>=E_n|\phi_n\>$) as
\beq
|r\>=\sum_nc_n|\phi_n\>.
\eeq
Here, $\{c_n\}$ are sampled randomly under the condition $\sum_n|c_n|^2=1$.
Using $|\phi_n\>$, the state $|\Psi\>$ is written as
\beq
|\Psi\>=\frac{1}{\mathcal{N}^{1/2}}\sqrt{\frac{D}{Z}}\sum_ne^{-\beta E_n/2}c_n|\phi_n\>,
\eeq
where $D$ is the dimension of the Hilbert space, $Z:=\sum_ne^{-\beta E_n}$ is the partition function, and
\beq
\mathcal{N}:=\frac{D}{Z}\sum_ne^{-\beta E_n}|c_n|^2.
\eeq

The average of $A$ over the random vectors is denoted by $\mathrm{Ave}[A]$.
It is shown that $\mathrm{Ave}[\mathcal{N}]=1$ and the probability of getting $\mathcal{N}$ with $|\mathcal{N}-1|\geq\epsilon (>0)$ is exponentially small.
More precisely, in Ref.~\cite{Sugiura-Shimizu2013}, it is shown that
\beq
\mathrm{Ave}\left[\chi(|\mathcal{N}-1|\geq\epsilon)\right]
\leq \frac{1}{\epsilon^2}e^{-2L\beta[f(2\beta)-f(\beta)]}=:\frac{1}{\epsilon^2}e^{-\eta L},
\label{eq:outside}
\eeq
where $\chi(\cdot)$ is the indicator function [$\chi(\mathrm{True})=1$ and $\chi(\mathrm{False})=0$], and $f(\beta):=-(1/\beta L)\ln Z$ is the free energy density.
Since $f(2\beta)>f(\beta)$ for $\beta>0$, we find $\eta>0$.

Now we consider the average weight to the nonthermal energy eigenstates, which is given by
\beq
\mathrm{Ave}[w]=\mathrm{Ave}[\<\Psi|\hat{\mathcal{Q}}|\Psi\>]
\eeq
with the projection operator $\hat{\mathcal{Q}}$ to the Hilbert subspace $\mathcal{T}$.
This quantity is decomposed as
\begin{align}
&\mathrm{Ave}[\<\Psi|\hat{\mathcal{Q}}|\Psi\>] \nonumber \\
&=\mathrm{Ave}\left[\chi(|\mathcal{N}-1|<\epsilon)\frac{1}{\mathcal{N}}\frac{D}{Z}\sum_ne^{-\beta E_n}|c_n|^2\<\phi_n|\hat{\mathcal{Q}}|\phi_n\>\right]
\nonumber \\
&+\mathrm{Ave}\left[\chi(|\mathcal{N}-1|\geq\epsilon)\frac{1}{\mathcal{N}}\frac{D}{Z}\sum_ne^{-\beta E_n}|c_n|^2\<\phi_n|\hat{\mathcal{Q}}|\phi_n\>\right]
\nonumber \\
&=: A_1+A_2,
\label{eq:decomp}
\end{align}
where $\epsilon$ is an arbitrary constant satisfying $0<\epsilon<1$.
We first calculate $A_1$.
Using $1/\mathcal{N}\leq 1/(1-\epsilon)$, $\<\phi_n|\hat{\mathcal{Q}}|\phi_n\>\leq 1$, and the Schwartz inequality, we obtain
\begin{align}
&A_1
\leq\frac{1}{1-\epsilon}\frac{D}{Z} \nonumber \\
&\times\mathrm{Ave}\left[\sqrt{\sum_n|c_n|^4e^{-\beta E_n}}\cdot\sqrt{\sum_n e^{-\beta E_n}\<\phi_n|\hat{\mathcal{Q}}|\phi_n\>}\right]
\nonumber \\
&\leq \frac{1}{1-\epsilon}\frac{D}{\sqrt{Z}}\sqrt{\<\hat{\mathcal{Q}}\>_{\mathrm{can}}^{(\mathrm{ini})}}\mathrm{Ave}\left[\sqrt{\sum_n|c_n|^4e^{-\beta E_n}}\right], 
\end{align}
where $\<\hat{A}\>_{\mathrm{can}}^{(\mathrm{ini})} :=\mathrm{Tr}\, e^{-\beta\hat{H}_{\mathrm{ini}}}\hat{A}/\mathrm{Tr}\, e^{-\beta\hat{H}_{\mathrm{ini}}}$ is the expectation value of $\hat{A}$ in the canonical ensemble of $\hat{H}_{\mathrm{ini}}$.
Since $\mathrm{Ave}[\sqrt{\cdot}]\leq\sqrt{\mathrm{Ave}[\cdot]}$,
\beq
\mathrm{Ave}\left[\sqrt{\sum_n|c_n|^4e^{-\beta E_n}}\right]
\leq \sqrt{\sum_n\mathrm{Ave}[|c_n|^4]e^{-\beta E_n}}.
\eeq
By using $\mathrm{Ave}[|c_n|^4]=2/D(D+1)\leq 2/D^2$, we finally obtain the upper bound of the first term of the right hand side of Eq.~(\ref{eq:decomp}):
\beq
A_1\leq \frac{1}{1-\epsilon}\sqrt{2\<\hat{\mathcal{Q}}\>_{\mathrm{can}}^{(\mathrm{ini})}}.
\label{eq:A1_ineq}
\eeq

As mentioned in Sec.~\ref{sec:thermalization}, one can prove that there exists some constant $\gamma>0$ independent of the system size $L$ such that
\beq
\<\hat{\mathcal{Q}}\>_{\mathrm{can}}^{(\mathrm{ini})}\leq e^{-\gamma L}.
\eeq
Using this in Eq.~(\ref{eq:A1_ineq}), we obtain
\beq
A_1\leq\frac{\sqrt{2}}{1-\epsilon}e^{-\gamma L/2}.
\label{eq:A1}
\eeq

Next, we consider $A_2$.
Using
\begin{align}
\frac{1}{\mathcal{N}}\frac{D}{Z}\sum_ne^{-\beta E_n}|c_n|^2\<\phi_n|\hat{\mathcal{Q}}|\phi_n\> \nonumber \\
=\frac{\sum_ne^{-\beta E_n}|c_n|^2\<\phi_n|\hat{\mathcal{Q}}|\phi_n\>}{\sum_ne^{-\beta E_n}|c_n|^2}\leq 1
\end{align}
and Eq.~(\ref{eq:outside}), we obtain
\beq
A_2\leq\mathrm{Ave}\left[\chi(|\mathcal{N}-1|\geq\epsilon)\right]\leq\frac{1}{\epsilon^2}e^{-\eta L}.
\label{eq:A2}
\eeq

Substituting Eqs.~(\ref{eq:A1}) and (\ref{eq:A2}) into Eq.~(\ref{eq:decomp}), we can conclude that $\mathrm{Ave}[\<\Psi|\hat{\mathcal{Q}}|\Psi\>]$ is exponentially small
\beq
\mathrm{Ave}[\<\Psi|\hat{\mathcal{Q}}|\Psi\>]\leq \( \frac{\sqrt{2}}{1-\epsilon}+\frac{1}{\epsilon^2}\) e^{-2\gamma'L}
\eeq
with
\beq
2\gamma'=\min\left\{\frac{\gamma}{2},\eta\right\}.
\eeq
By applying the Markov inequality~\cite{Feller_text}, we obtain
\begin{align}
\mathrm{Prob}\left[\<\Psi|\hat{\mathcal{Q}}|\Psi\>\geq e^{-\gamma'L}\right]\leq e^{\gamma' L}\mathrm{Ave}[\<\Psi|\hat{\mathcal{Q}}|\Psi\>]
\nonumber \\
\leq \( \frac{\sqrt{2}}{1-\epsilon}+\frac{1}{\epsilon^2}\) e^{-\gamma'L},
\end{align}
where $\mathrm{Prob}[a]$ is the probability of an event $a$.
Here it is noted that we can choose $\epsilon$ as an arbitrary positive value less than 1, for example, we put $\epsilon=1/2$, and then we obtain Eq.~(\ref{eq:prob_weight}).

\section{Occurrence of prethermalization}
\label{app:occurrence}

We shall derive prethermalization in the spin-1 model discussed in the main text.
Let us consider the time evolution of a local operator $\hat{O}_X$ that acts non-trivially on the set of sites $X=\{j,j+1,\dots,j+k-1\}$ with some integer $k\geq 1$ independent of the length $L$ of the one-dimensional chain.
For some fixed $l\geq 0$, we define the region $S$ as a set of sites $i$ with $\min_{j'\in X}\mathrm{dist}(i,j')\leq l$, where $\mathrm{dist}(i,j')$ is the distance between the sites $i$ and $j'$ in the periodic boundary condition.
Explicitly, $S$ is given by 
\beq
S=\{j-l,j-l+1,\dots,j+k+l-2,j+k+l-1\}.
\eeq
The set of sites not in $S$ is denoted by $B$.

We divide the Hamiltonian $\hat{H}$ as
\beq
\hat{H}=\hat{H}_S+\hat{H}_B+\hat{H}_{SB},
\eeq
where $\hat{H}_S$ is the Hamiltonian acting non-trivially on the region $S$, $\hat{H}_B$ is that on the region $B$, and $\hat{H}_{SB}$ is the interaction Hamiltonian between the subsystems $S$ and $B$.
Since $\hat{H}$ is a local Hamiltonian, we can always choose $\hat{H}_{SB}$ so that $\|\hat{H}_{SB}\|$ is independent of $L$, where $\|\cdot\|$ represents the operator norm.
The whole Hilbert space $\mathcal{H}$ is accordingly decomposed as $\mathcal{H}=\mathcal{H}_S\otimes\mathcal{H}_B$.
In the spin-1 Hamiltonian discussed in the main text, the concrete expressions are given as follows:
\beq
\left\{
\begin{split}
\hat{H}_S&=\sum_{i=j-l+1}^{j+k+l-2}\hat{h}_i\hat{P}_i+\sum_{i=j-l}^{j+k+l-2}\hat{h}_i'=:\hat{H}_{0S}+\hat{H}_S',
\\
\hat{H}_{SB}&=\hat{h}_{j-l-1}\hat{P}_{j-l-1}+\hat{h}_{j-l}\hat{P}_{j-l}
\\
&+\hat{h}_{j+k+l-1}\hat{P}_{j+k+l-1}+\hat{h}_{j+k+l}\hat{P}_{j+k+l}
\\
&+\hat{h}_{j-l-1}'+\hat{h}_{j+k+l-1}',
\end{split}
\right.
\eeq
where $\hat{H}'=\sum_{i=1}^L\hat{h}_i'$ with
\beq
\hat{h}_i'=\sum_{\alpha=x,y,z}(J_{\alpha}'\tilde{\sigma}_i^{\alpha}\tilde{\sigma}_{i+1}^{\alpha}-h_{\alpha}'\tilde{\sigma}_i^{\alpha}).
\eeq
The Hamiltonian $\hat{H}_{SB}$ acts non-trivially on the set of sites
\begin{align}
\d S:=\{j-l-2,j-l-1,j-l,j-l+1,
\nonumber \\
j+k+l-2,j+k+l-1,
\nonumber \\
j+k+l,j+k+l+1\}.
\end{align}

We shall show that the exact time evolution of $\hat{O}_X$ is well approximated by its time evolution under $\hat{H}_S$, that is,
\beq
e^{i\hat{H}t}\hat{O}_Xe^{-i\hat{H}t}\approx e^{i\hat{H}_St}\hat{O}_Xe^{-i\hat{H}_St}
\label{eq:LR_approx}
\eeq
up to some finite time.
Using the identity $e^{i(\hat{A}+\hat{B})t}=e^{i\hat{A}t}+i\int_0^tds e^{i(\hat{A}+\hat{B})(t-s)}\hat{B}e^{i\hat{A}s}$ for the super-operators $\hat{A}=[\hat{H}_S,\cdot]$ and $\hat{B}=[\hat{H}_B+\hat{H}_{SB},\cdot]$, we obtain
\begin{align}
&\left\|e^{i\hat{H}t}\hat{O}_Xe^{-i\hat{H}t}-e^{i\hat{H}_St}\hat{O}_Xe^{-i\hat{H}_St}\right\|
\nonumber \\
&=\left\|\int_0^tdse^{i\hat{H}(t-s)}[H_B+H_{SB},e^{i\hat{H}_Ss}\hat{O}_Xe^{-i\hat{H}_Ss}]e^{-i\hat{H}(t-s)}\right\|
\nonumber \\
&\leq\int_0^tds\left\|[\hat{H}_{SB},e^{i\hat{H}_Ss}\hat{O}_Xe^{-i\hat{H}_Ss}]\right\|.
\label{eq:LR_diff}
\end{align}
Using the Lieb-Robinson bound~\cite{Lieb1972,Hastings2006}, we have
\begin{align}
&\left\|[\hat{H}_{SB},e^{i\hat{H}_Ss}\hat{O}_Xe^{-i\hat{H}_Ss}]\right\|
\nonumber \\
&\leq C|\d S|\cdot|X|\cdot\|\hat{H}_{SB}\|\cdot\|\hat{O}_X\|e^{-\mu\mathrm{dist}(\d S,X)}e^{v|s|}
\nonumber \\
&=8Ck\|\hat{H}_{SB}\|\|\hat{O}_X\|e^{-\mu(l-1)}e^{v|s|},
\label{eq:LR}
\end{align}
where we used $|\d S|=8$, $|X|=k$, and $\mathrm{dist}(\d S,X):=\min_{i'\in\d S,j'\in X}\mathrm{dist}(i',j')=l-1$ in this model.
$C$ is a positive constant independent of the system size; in our model, for example, we can set $C=2/5$.
$\mu$ and $g$ are arbitrary positive constants satisfying
\beq
\sum_{n=1}^{3}\left\|\hat{h}_{n}\hat{P}_{n}+\hat{h}_{n}'\right\|\leq g e^{-2\mu},
\eeq
and $v=4g/C$.
See Ref.~\cite{Hastings2006} for details.

Substituting Eq.~(\ref{eq:LR}) into Eq.~(\ref{eq:LR_diff}) with $C':=8Cke^{\mu}\|\hat{H}_{SB}\|/v$, we arrive at
\begin{align}
\left\|e^{i\hat{H}t}\hat{O}_Xe^{-i\hat{H}t}-e^{i\hat{H}_St}\hat{O}_Xe^{-i\hat{H}_St}\right\|
\nonumber \\
\leq C'\|\hat{O}_X\|e^{-\mu l+vt}
=C'\|\hat{O}_X\|e^{-\mu(l-v_{\mathrm{LR}}t)},
\label{eq:LR_bound}
\end{align}
where $v_{\mathrm{LR}}:=v/\mu$ is called the Lieb-Robinson velocity.
This inequality implies that the approximation of Eq.~(\ref{eq:LR_approx}) is valid for $t\lesssim l/v_{\mathrm{LR}}=:\tau_{\mathrm{LR}}^S$, which is called the Lieb-Robinson time.

After the approximation given by Eq.~(\ref{eq:LR_approx}), the time evolution is fully determined by the Hamiltonian of the finite subsystem $S$.
Although the weight $w$ to the nonthermal energy eigenstates of $\hat{H}$ is exponentially small with respect to the total system size $L$, the weight $w_S$ to the nonthermal energy eigenstates of 
$\hat{H}_S$ is not necessarily small.
According to the discussion in the main text, under the time evolution in $\hat{H}_S$, the subsystem will relax to the stationary state described by the nonlocal GGE given by
\beq
\rho_{\mathrm{GGE}}^{S}=\frac{1}{Z_{\mathrm{GGE}}}e^{-\beta_P\hat{\mathcal{P}}_S\hat{H}_S\hat{\mathcal{P}}_S -\beta_Q\hat{\mathcal{Q}}_S\hat{H}_S\hat{\mathcal{Q}}_S-\lambda\hat{\mathcal{Q}}_S},
\eeq
where $\hat{\mathcal{Q}}_S:=\prod_{i\in S}\hat{Q}_i$ and $\hat{\mathcal{P}}_S:=1-\hat{\mathcal{Q}}_S$.
Therefore, if the relaxation time $\tau_S$ under the time evolution by $\hat{H}_S$ is much shorter than the Lieb-Robinson time, $O_X^S(t):=\<\Psi(0)|e^{i\hat{H}_St}\hat{O}_Xe^{-i\hat{H}_St}|\Psi(0)\>$ will relax to a prethermalized value given by $\mathrm{Tr}\,\rho_{\mathrm{GGE}}^S\hat{O}_X$ when $\tau_S\ll t\lesssim\tau_{\mathrm{LR}}^S$.

The condition $\tau_S\ll\tau_{\mathrm{LR}}^S$ is generally not met since the relaxation of $S$ takes place due to the spread of defects over the entire region of $S$, the timescale of which cannot be shorter than the Lieb-Robinson time.
Only when $w_S\approx 1$, the condition $\tau_S\ll\tau_{\mathrm{LR}}^S$ may be satisfied, as we see below.

Let us consider the situation of $w_S\approx 1$.
This situation is realized when
the initial density of the defects
\beq
d_0=\frac{1}{L}\sum_{i=1}^L\<\Psi(0)|\hat{P}_i|\Psi(0)\>
\eeq
is very small.
When the defects are distributed uniformly over the entire system,
\beq
w_S\approx (1-d_0)^{|S|}=(1-d_0)^{k+2l}\leq 1-(k+2l)d_0.
\eeq
We assume that $d_0k\ll 1$.
In order for $w_S$ to be close to 1,
\beq
l\ll \frac{1}{d_0},
\eeq
and hence
\beq
\tau_{\mathrm{LR}}^S\ll \frac{1}{v_{\mathrm{LR}}d_0}.
\label{eq:tau_LR}
\eeq
When $w_S\approx 1$, the relaxation time of the subsystem $S$ is independent of $d_0$ because the relaxation takes place within the subspace $\mathcal{T}_S$.
Moreover, if the initial state in the subsystem $S$ is homogeneous (no inhomogeneity of the energy $\hat{H}_S$), the relaxation time of the subsystem $S$ is also independent of $k+2l$, i.e., the size of the subsystem $S$, for large $l$.
Therefore, for sufficiently small $d_0$, we can choose $l$ so that
\beq
\tau_S\ll\tau_{\mathrm{LR}}^S\ll \frac{1}{v_{\mathrm{LR}}d_0},
\label{eq:l}
\eeq
and then both $w_S\approx 1$ and $\tau_S\ll\tau_{\mathrm{LR}}^S$ can be realized at the same time.
This implies that prethermalization occurs in this situation.

Since $w_S=\<\Psi(0)|\hat{\mathcal{Q}}_S|\Psi(0)\>$, $w_S\approx 1$ implies
\beq
\rho_{\mathrm{GGE}}^S\approx\frac{\hat{\mathcal{Q}}_Se^{-\beta_Q\hat{\mathcal{Q}}_S\hat{H}_S\hat{\mathcal{Q}}_S}}{\mathrm{Tr}_S\,\hat{\mathcal{Q}}_Se^{-\beta_Q\hat{\mathcal{Q}}_S\hat{H}_S\hat{\mathcal{Q}}_S}}=:\rho_Q^S,
\label{eq:nonlocal_GGE_S}
\eeq
which is nothing but the constrained equilibrium to the subspace $\mathcal{T}_S$, where $\mathcal{T}_S$ is defined as the set of states $|\Psi\>\in\mathcal{H}_S$ satisfying $\<\Psi|\hat{\mathcal{Q}}_S|\Psi\>=1$, i.e., $\mathcal{T}_S$ is the subspace of the nonthermal energy eigenstates of $\hat{H}_S$.
When $d_0$ is sufficiently small, we can choose a large value of $l$ satisfying Eq.~(\ref{eq:l}), which implies that the size of the region $S$ is sufficiently large.
In this case, the density matrix (\ref{eq:nonlocal_GGE_S}) is approximately identical to the reduced density matrix obtained from the constrained equilibrium density matrix of the whole system, that is,
\beq
\left\{
\begin{split}
&\rho_Q^S\approx\mathrm{Tr}_B\,\rho_Q, \\
&\rho_Q=\frac{\hat{\mathcal{Q}}e^{-\beta_Q\hat{\mathcal{Q}}\hat{H}\hat{\mathcal{Q}}}}{\Tr\, \hat{\mathcal{Q}}e^{-\beta_Q\hat{\mathcal{Q}}\hat{H}\hat{\mathcal{Q}}}}.
\end{split}
\right.
\label{eq:nonlocal_GGE_app}
\eeq
Equation (\ref{eq:nonlocal_GGE_app}) tells us that the prethermalized plateau of $\<\Psi(t)|\hat{O}_X|\Psi(t)\>$ is described by the constrained equilibrium to $\mathcal{T}$ of the whole system.

We shall evaluate the timescale of prethermalization.
The condition of Eq.~(\ref{eq:tau_LR}) implies that at time given by
\beq
t_{\mathrm{pre}}^{\mathrm{(LR)}}\sim\frac{1}{v_{\mathrm{LR}}d_0},
\eeq
the prethermalized state will decay towards the true thermal equilibrium.
This gives a lower bound of the timescale of prethermalization.

\end{document}